\documentclass{article}

\usepackage{hyperref}

\usepackage[utf8]{inputenc} 
\usepackage[T1]{fontenc}    
\usepackage{hyperref}       
\usepackage{url}            
\usepackage{booktabs}       
\usepackage{amsfonts}       
\usepackage{nicefrac}       
\usepackage{microtype}      
\usepackage{lipsum}
\usepackage{graphicx}
\usepackage[rightcaption]{sidecap}
\usepackage{float}
\usepackage[all]{nowidow}

\usepackage{amsmath}

\begin{document}


\title{Reliability and comparability of human brain structural covariance networks}

\author{Jona Carmon$^{1}$, Jil Heege$^{2}$, Joe H Necus$^{3}$, Thomas W Owen$^{3}$, \\Gordon Pipa$^{1}$, Marcus Kaiser$^{3,4,5}$, Peter N Taylor$^{3,4,6}$, \\Yujiang Wang$^{*3,4,6}$}

\maketitle 

\begin{enumerate}
\item{Institute of Cognitive Science, Osnabrueck University, Osnabrueck, Germany}
\item{Humboldt University Berlin, Berlin, Germany}
\item{CNNP Lab (www.cnnp-lab.com), Interdisciplinary Computing and Complex BioSystems Group, School of Computing, Newcastle University, Newcastle upon Tyne, United Kingdom}
\item{Faculty of Medical Sciences, Newcastle University, Newcastle upon Tyne, United Kingdom}
\item{School of Medicine, Shanghai Jiao Tong University, Shanghai, China}
\item{UCL Queen Square Institute of Neurology, Queen Square, London, United Kingdom}
\end{enumerate}
* Yujiang.Wang@newcastle.ac.uk
\newpage

\begin{abstract}
Structural covariance analysis is a widely used structural MRI analysis method which characterises the co-relations of morphology between brain regions over a group of subjects. To our knowledge, little has been investigated in terms of the comparability of results between different data sets of healthy human subjects, as well as the reliability of results over the same subjects in different rescan sessions, image resolutions, or FreeSurfer versions. 

In terms of comparability, our results show substantial differences in the structural covariance matrix between data sets of age- and sex-matched healthy human adults. These differences persist after univariate site correction, they are exacerbated by low sample sizes, and they are most pronounced when using average cortical thickness as a morphological measure. Down-stream graph theoretic analyses further show statistically significant differences. 

In terms of reliability, substantial differences were also found when comparing repeated scan sessions of the same subjects, image resolutions, and even FreeSurfer versions of the same image. We could further estimate the relative measurement error and showed that it is largest when using cortical thickness as a morphological measure.  Using simulated data, we argue that cortical thickness is least reliable because of larger relative measurement errors.

Practically, we make the following recommendations (1) combining subjects across sites into one group should be avoided, particularly if sites differ in image resolutions, subject demographics, or preprocessing steps; (2) surface area and volume should be preferred as morphological measures over cortical thickness; (3) a large number of subjects ($n>>30$ for the Desikan-Killiany parcellation) should be used to estimate structural covariance; (4) measurement error should be assessed where repeated measurements are available; (5) if combining sites is critical, univariate (per ROI) site-correction is insufficient, but error covariance (between ROIs) should be explicitly measured and modelled.

\end{abstract}

\section{Introduction}

Different brain regions have distinct morphologies, and these morphological differences co-vary between brain regions within the human population \cite{mechelli2005structural}. This covariance has been related to functional connectivity, genetics, and alteration in disease \cite{alexander2013imaging}, although the biological interpretation of this covariance remains under discussion. In the field of `structural covariance analysis', the covariance between brain regions is measured. 

\begin{figure}
  \centering
  \hspace*{-1cm}
  \includegraphics[width=140mm]{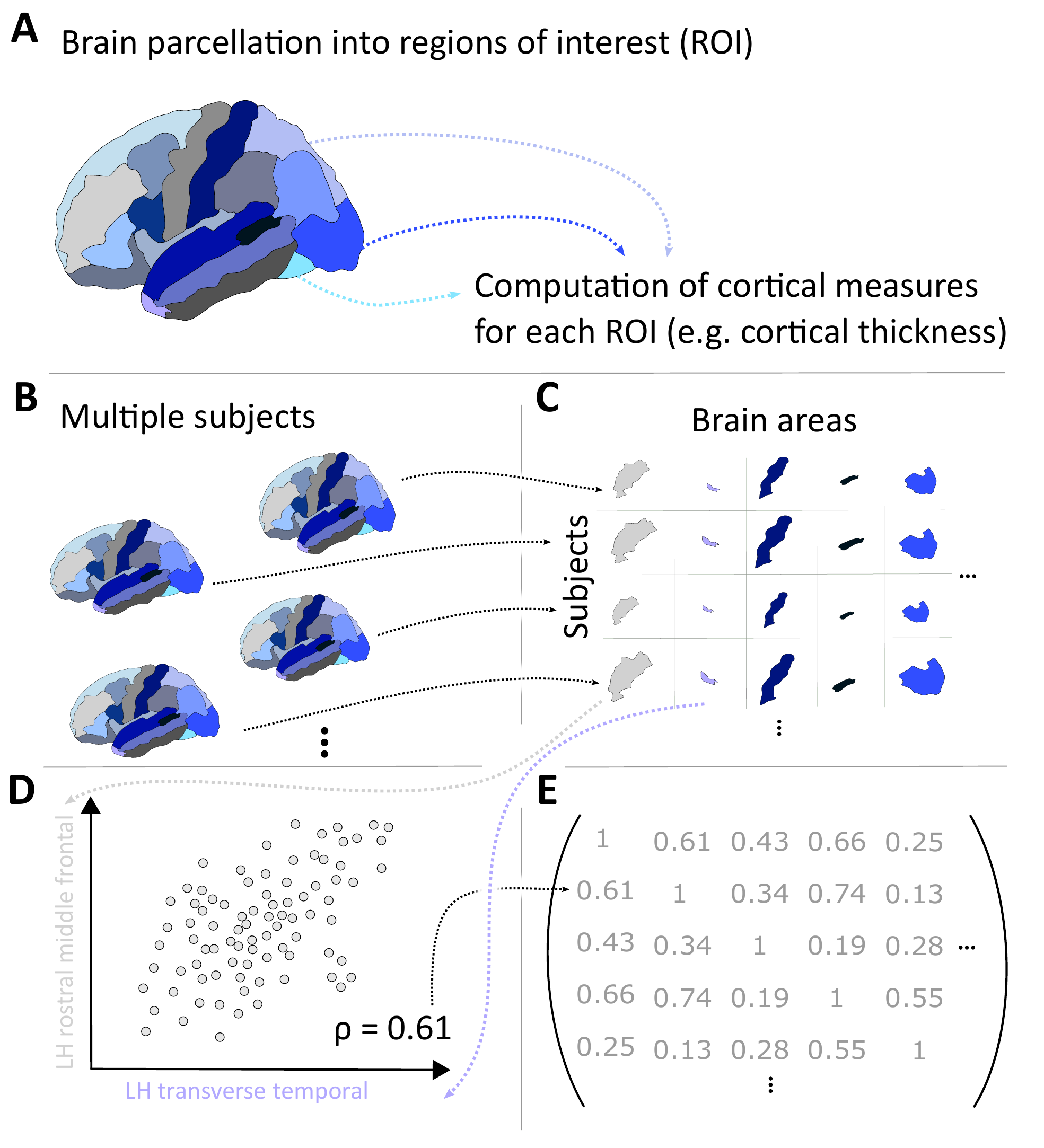}
  \caption{\textbf{Schematic illustration of structural covariance analysis.} (A) Choice of brain atlas and computation of cortical measures. (B) Multiple subjects in one data set. (C) Construction of matrix with brain region and subjects. Typically at this point one corrects for the subjects age and sex. (D) Computation of correlation between brain regions. (E) The matrix of all correlations is termed the structural covariance.}
  \label{StructCov}
\end{figure}

Typically, structural covariance is computed from structural (T1 weighted) Magnetic Resonance Imaging (MRI) data, from which cortical morphology measures (e.g. cortical thickness, surface area, and volume) are inferred using software tools such as FreeSurfer. The cortical morphology measures are computed for every brain region of a chosen brain atlas (Fig.~\ref{StructCov}(A)) and every subject of the selected data set (Fig.~\ref{StructCov}(B)). These estimates can be captured in a matrix of subjects as rows and brain areas as columns (Fig.~\ref{StructCov}(C)). From this matrix the correlation between every pair of brain region is computed, (Fig.~\ref{StructCov}(D)), and stored as a `structural covariance matrix' which captures the correlations between all pairs of brain regions (Fig.~\ref{StructCov}(E)).

Neuroimaging research increasingly analyses structural covariance. It is, for example, applied to compare brain networks between healthy subjects and different clinical conditions including Schizophrenia \cite{moberget2018cerebellar, bassett2008hierarchical, mitelman2005cortical, bhojraj2010inter, spreng2019structural}, Alzheimer's disease \cite{hafkemeijer2016differences, li2019gray, yao2010abnormal, desikan2010selective}, Autism \cite{bethlehem2017structural,mcalonan2004mapping} or Epilepsy \cite{bernhardt2011graph}. For such clinical studies, the wide availability and high quality of T1 weighted MRI scans is advantageous compared to other modalities. Many studies using structural covariance analyses also combine data sets from multiple sites \cite{bethlehem2017structural,moberget2018cerebellar, hafkemeijer2016differences,li2019gray, desikan2010selective}. With the increasing availability of public data sets, more structural covariance analysis of pooled data sets are to be expected. 

Many previous studies have examined the reliability of using quantitative measures of cortical morphology \cite{dickerson2008detection, iscan2015test, gronenschild2012effects, han2006reliability}. In our study we refer to `reliability' as the quantitative consistency of different measurements of the same subjects. Furthermore, we will use the term `comparability' to describe how consistent measurements are between data sets of different subjects with comparable demographics (e.g. same age and sex). Dickerson \textit{et al.} investigated the reliability of computationally inferred cortical measures for scan sessions, scanners and field strength \cite{dickerson2008detection}. They conclude that the tested automated measures of cortical thickness are highly reliable within scanner systems and across manufactures and field strength \cite{dickerson2008detection}. Iscan \textit{et al.} investigated the comparability and reliability of FreeSurfer estimates within and between sites and found that the comparability of average cortical thickness is more sensitive to site differences than volume or surface area \cite{iscan2015test}. Grovenschild \textit{et al.} investigated the reliability of FreeSurfer estimates between different FreeSurfer versions. They found that the FreeSurfer version can have a strong effect on the estimates of the cortical morphology measures. The differences they found were on average 1.3-64\% for volume and 1.1-7.7\% for average cortical thickness \cite{gronenschild2012effects}. However, despite a number of studies on reliability and comparability of the raw cortical morphology measures, little has been investigated on the reliability and comparability of structural covariance. The comparability aspect is of especially high relevance for studies that combine multisite data for their analysis.

To investigate comparability in this study, we compared the structural covariance for data sets of healthy human subjects with comparable demographics. For our analysis of reliability, we compared the structural covariance of the same subjects between rescan sessions, and for the same scans the reliability of different FreeSurfer versions. Additional investigations of reliability and comparability with a range of other datasets are presented in the supplementary materials. Finally, we investigated if, and why, different cortical morphology measures are differentially comparable and reliable for structural covariance analyses. 

\section{Methods}
\subsection{Data selection and pre-processing}
To determine the comparability and reliability of structural covariances we analysed two data sets in the main text and additional dataset in the supplementary materials. The two datasets used in the main results are: The Human Connectome Project (HCP), which can be found under \url{https://db.humanconnectome.org/} \cite{van2012human}, and the Cambridge Centre for Ageing and Neuroscience (Cam-CAN) data is available at \url{http://www.mrc-cbu.cam.ac.uk/datasets/camcan/} \cite{shafto2014cambridge,taylor2017cambridge}. The MRI scanners of both data sets have 3T field strength. HCP uses a customized Siemens Skyra scanner with 0.7~mm isotropic voxel size located at Washington University, St. Louis, MO, USA. Cam-CAN uses a Siemens TIM Trio System with 1~mm isotropic voxel size located at the Medical Research Council Cognition and Brain Sciences Unit, Cambridge, UK. 



We focus on three main comparisons of structural covariance in the main text: comparing sites, FreeSurfer processing versions, and scan/rescan effects. The supplementary materials provide additional comparisons in terms of other factors or datasets.

For the site comparison, we intentionally selected settings that differ slightly between the two sites to be compared: (different image resolution and FreeSurfer version) to simulate a realistic scenario that a multi-site study may encounter. To this end, we selected 100 unrelated subjects from the HCP data set and used the preprocessed FreeSurfer folders provided by HCP (using a modified version of FreeSurfer 5.3) \cite{glasser2013minimal}. The original image resolution for HCP is 0.7mm isotropic. The structural MRI images of Cam-CAN were pre-processed with FreeSurfer versions 6.0 using the `recon-all' script for segmentation, surface reconstruction and parcellation. The original image resolution for CamCAN is 1mm isotropic. We retained 644 subjects that successfully completed recon-all without errors. We removed one subject from the HCP data as an outlier (subject ID: 414229). To reduce confounding effects of age and sex, the subjects were age-and-sex matched: we selected 43 males and 43 females in the age range of 22-34 from both data sets (age 28.26 $\pm$ 3.53 in Cam-CAN and 28.7 $\pm$ 3.49 in HCP). In one part of our analysis (Fig. \ref{NetworkDiff}), we also investigated if reducing some of the image resolution and pre-processing differences would render sites more comparable in terms of their structural covariance. To this end, we also created a downsampled HCP version (to 1mm isotropic using  interpolation in FreeSurfer mri\_convert), which we processed using FreeSurfer 5.3. We further also processed the CamCAN data using FreeSurfer 5.3. Finally, we provide additional site comparisons in Supplementary Materials S16 and S17.

For the analysis of reliability of structural covariances between different FreeSurfer versions, we used the above-mentioned Cam-CAN dataset pre-processed in FreeSurfer version 5.3 and 6.0. We selected the same 86 subjects as before for the comparison of sites. 

To compare structural covariance between scan/rescan sessions, we used the HCP scan and rescan data set. The data set comprises 45 subjects who were scanned at two different scanning sessions typically no more than two years apart. Again, we used the preprocessed FreeSurfer folders provided by HCP. The Supplementary Text S14 includes an additional comparison of scan sessions in another dataset with a shorter inter-scan interval, which also agrees and further supports our main findings.

Table~\ref{tableoverview} provides an overview of all the subject numbers and demographics used in our main analysis.


 \begin{table}
 \scriptsize
\begin{tabular}{ |p{2.5cm}||p{2cm}|p{2cm}|p{2.5cm}| }
 \hline
 &&&\\
 \textbf{Data set name} & \textbf{Number of subjects} &\textbf{Number of females} &\textbf{Age range} \\
 &&&\\
 \hline
  &&&\\
  \textbf{Site comparison:}&&&\\
 Cam-CAN &   86  & 43 &28.26 $\pm$ 3.53\\
 HCP   & 86    & 43 & 28.7 $\pm$ 3.49 \\
  &&&\\
  \hline
  &&&\\
    \textbf{FreeSurfer version comparison:}&&&\\
 Cam-CAN FreeSurfer 6.0 & 86  & 43 &28.26 $\pm$ 3.53 \\
 Cam-CAN FreeSurfer 5.3   & 86  & 43 &28.26 $\pm$ 3.53 \\
  &&&\\
  \hline
  &&&\\
\textbf{Scan session comparison:}&&&\\
 HCP Scan & 45 &  31&  range: 22-35 \\
 HCP Rescan   & 45 &  31&  range: 22-35 \\
  &&&\\
 \hline
\end{tabular}
\caption{\textbf{Overview table of subjects used for different comparisons in the main text.} We provide the age range as mean $\pm$ standard deviation to indicate how closely subject groups have been matched between HCP and CamCAN. For the HCP scan and rescan data we used the entire data set provided.} \label{tableoverview}
\end{table}

To compute the structural covariance matrix, we used the Desikan Killiany parcellation \cite{desikan2006automated} in FreeSurfer throughout, if not otherwise stated. The Desikan Killiany parcellation comprises 34 brain regions in each hemisphere, and is commonly chosen in structural covariance analysis.
In one subsequent analysis (Fig.~\ref{SingleCorrError}), to provide a simplified illustration, we also correlated the left and right hemisphere across subjects to obtain a single correlation coefficient (instead of a correlation matrix). To this end, we obtained the average cortical thickness estimates of the left and the right hemisphere directly from FreeSurfer. For volume and surface area, we summed the ROIs of the Desikan Killiany atlas to compute the estimates for the left and the right hemisphere.

Finally, Supplementary Text S15 provides additional comparisons of structural covariance derived from different image resolutions.

\subsection{Data analysis and visualisation}

\subsubsection{Statistical analysis\label{sectionstats}}

We standardised each measure across all subjects, if not explicitly stated otherwise, which normalizes the mean of each ROI to 0 and the standard deviation of each ROI to 1. This step has no effect on the calculation of correlations, but permits later investigation of the relative error variance (relative to the variance of 1 of the measurement).
To compute differences in correlation, we Fisher transformed the correlation values first. 

\subsubsection{Covariance ellipse for visualisation purposes \label{ellipsemethod}}
To visualise some correlations/covariances of our study, we display them as an ellipse. The ellipse represents the region that contains 95\% of all samples that can be drawn from the underlying covarying Gaussian distribution. We calculated the 95\% confidence interval with the Chi-squared distribution. The alignment of the error ellipse is computed upon the eigenvalues of the covariance matrix of the respective ROIs. The eigenvector with the largest eigenvalue determines the direction of the the ellipsoid longer axis. Since eigenvectors are orthogonal, the shorter axis of the ellipsoid corresponds to the direction of the smaller eigenvector. Our code was inspired by Matlab source code provided by \cite{spruyt2014draw, spruytweb}. To compute ellipsoids for the estimated error structure and estimated underlying correlation, we use our estimated variance and covariance (see section \ref{esterrorsection}).

\subsubsection{Network measures}
We computed several common network measures on the structural covariance matrix. As a first step, we thresholded (and binarised) the structural covariance matrix. We performed thresholding as a percentage of the network density. I.e. a threshold of 0.1, (10\%) indicates that the top 10\% of  strongest edges (in absolute value) are retained. We computed different brain network measures with the brain connectivity toolbox \cite{rubinov2010complex}. We investigate node strength (or node degree for the binarised matrices), characteristic path and clustering coefficient. These network measures are typically used in downstream analysis of structural covariance. In the supplementary material we additionally include the global efficiency, eigenvector centrality, assortativity and k-core centrality.

\subsubsection{Permutation test}

\begin{figure}[H]
  \centering
  \hspace*{-1.5cm}
  \includegraphics[width=140mm]{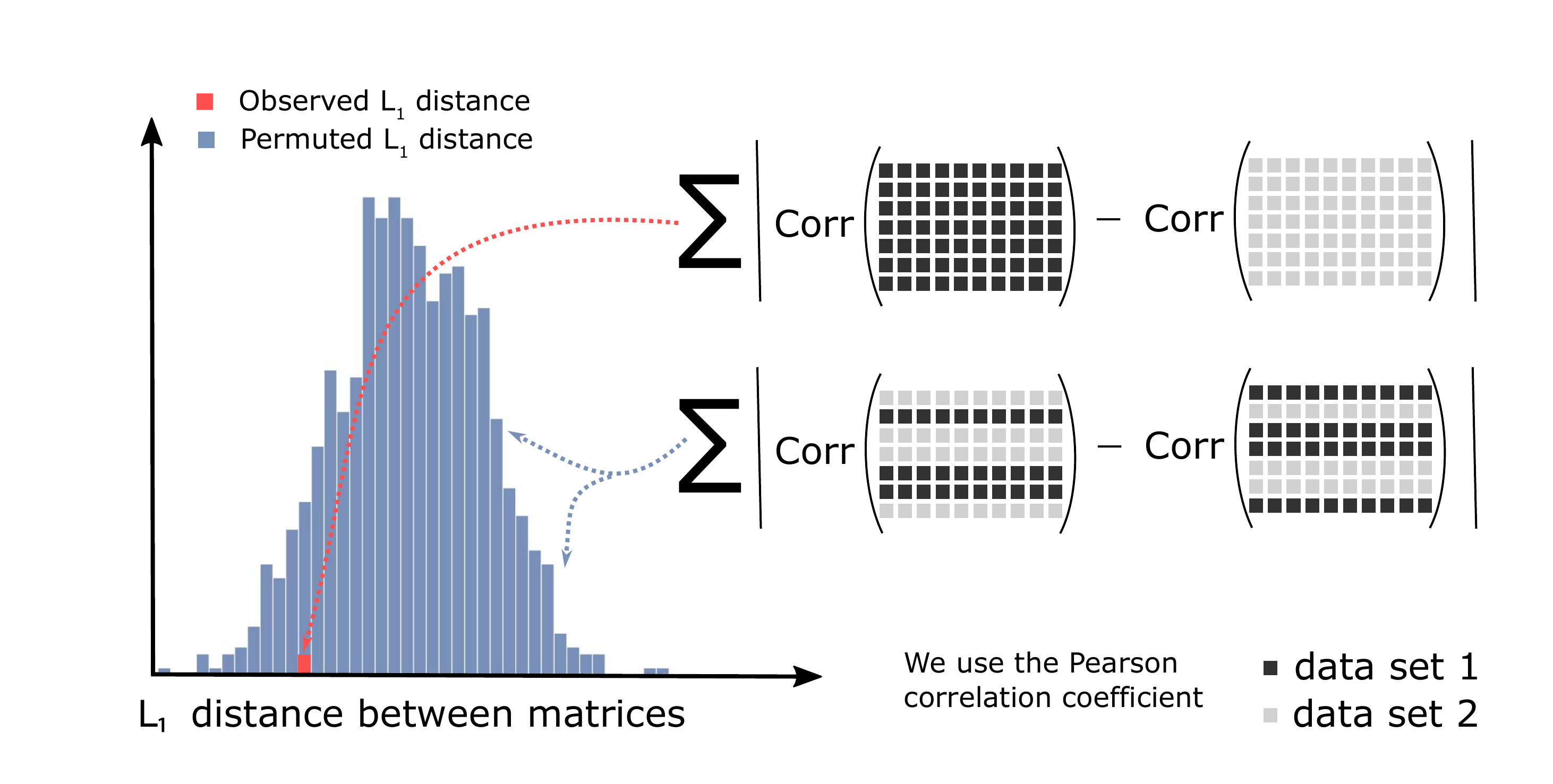}
  \caption{\textbf{Schematic illustration of the permutation test on the example of L1 distance between the structural covariance matrices.} We computed the $L_{1}$ distance (red marker) from the structural covariance matrices of the respective data sets. We repeated this process 1000 times with randomly mixed data splits to gain a distribution which serves as reference. With this distribution we can compute the p-value of the actual data split.}
  \label{PermutationTest}
\end{figure}

To compare the structural covariance matrices and network measures between data sets for statistical significant differences, we apply a permutation test (Fig.~\ref{PermutationTest}). E.g. to compute the difference of the two matrices, we calculated the $L_{1}$ distance (sum of the absolute differences). To obtain the reference distribution, we then computed the $L_{1}$ distance of 1000 randomly mixed data splits. The p-value of the actual difference is estimated from this distribution. Such permutation tests are common in studies that are e.g. comparing a patient vs. a control group. We demonstrate here the effect of such a test on our comparisons of, for example, two healthy subject data sets. We do not use the concept of statistical significance in further downstream analyses (e.g. for the selection of specific ROIs). 

\subsubsection{Estimation of error structure and underlying correlation \label{esterrorsection}}
To analyse reliability of repeated measurements of the same subjects, we estimated an error covariance structure of the data. Such repeated measurements were for example scan vs. rescan of the same subject, or applying FreeSurfer 5.3 vs. 6.0 in the same subject and scan. Here we explain how we estimated the error and the underlying correlation for the left and right hemisphere. The estimations for all other pairs of brain regions are analogous. We denote $LH_1 = X + E_1$ for the first measurement of the left hemisphere and  $LH_2 = X + E_2$ for the second measurement of the left hemisphere.  Respectively we denote $RH_1 = Y + D_1$ for the first measurement of the right hemisphere and $RH_2 = Y + D_2$ for the second measurement of the right hemisphere. $E_1,E_2, D_1$ and $D_2$ are errors introduced at this measurement. $X$ and $Y$ capture the actual biological measure of interest (in reality, this will also include systematic errors of this measurement as well as random errors of other processing steps). Note that $X$, $Y$, $E_1$,$E_2$,$D_1$ and $D_2$ are vectors of the same length, where each entry in the vectors corresponds to one subject.

For our derivation we make the following independence assumptions:
\newline
\newline
\begin{equation*}   
\begin{aligned}
X &\perp\!\!\!\perp E_1,E_2,D_1, D_2&\\
Y &\perp\!\!\!\perp E_1,E_2,D_1, D_2&\\
E_1 &\perp\!\!\!\perp E_2&\\
D_1 &\perp\!\!\!\perp D_2&\\
E_1 &\perp\!\!\!\perp D_2&\\
E_2 &\perp\!\!\!\perp D_1&\\
\end{aligned}
\end{equation*}
\newline

$Variable_1 \perp\!\!\!\perp Variable_2$ indicates that $Variable_1$ and $Variable_2$ are independent. In other words, we assume the actual measurements ($X$ and $Y$) are independent of the errors, and the errors between the repeat measurements are independent of each other.

\paragraph{Calculation of the variances} \mbox{}\\
\newline
With the following quantities which can be directly calculated from the empirically measured data:
\newline
\newline
\begin{equation*}
\begin{aligned}
&\alpha = \mathrm{Var}[(X + E_1) - (X +E_2)] = \mathrm{Var}[E_1] + \mathrm{Var}[E_2]\\
&\beta = \mathrm{Var}[X + E_1] = \mathrm{Var}[X] + \mathrm{Var}[E_1]\\
&\gamma = \mathrm{Var}[X + E2] = \mathrm{Var}[X] + \mathrm{Var}[E_2]\\
\end{aligned}
\end{equation*}
\newline
we then get
\newline
\newline
\begin{equation*}
\begin{aligned}
&\mathrm{Var}[E_1] = \frac{\alpha + \beta - \gamma}{2}\\
&\mathrm{Var}[E_2] = \frac{\alpha - \beta + \gamma}{2}\\
&\mathrm{Var}[X] = \frac{- \alpha + \beta + \gamma}{2}\\
\end{aligned}
\end{equation*}
\newline
Analogously to that we can compute $\mathrm{Var}[Y]$, $\mathrm{Var}[D_1]$ and $\mathrm{Var}[D_2]$.
\newline

\paragraph{Calculation of the covariances}\mbox{}\\

Similarly the following covariances can be directly calculated from the empirical data:
\newline
\newline
\begin{equation*}
\begin{aligned}
\delta &= \mathrm{Cov}[X+ E_1, Y + D_1] = \mathrm{Cov}[X,Y] + \mathrm{Cov}[E_1,D_1]\\
\epsilon &= \mathrm{Cov}[X + E_2, Y + D_2] = \mathrm{Cov}[X,Y] + \mathrm{Cov}[E_2,D_2]\\
\zeta &= \mathrm{Cov}[(X + E_1) - (X + E_2), (Y+D_1) - (Y + D_2)] \\
\quad\quad &= \mathrm{Cov}[E_1 - E_2, D_1 - D_2] =\mathrm{Cov}[E_1,D_1] + \mathrm{Cov}[E_2,D_2]\\
 \end{aligned}
\end{equation*}
\newline
we get
\newline
\newline
\begin{equation*}
\begin{aligned}
&\mathrm{Cov}[E_1,D_1] = \frac{\delta - \epsilon + \zeta}{2}\\
&\mathrm{Cov}[E_2,D_2] =  \frac{- \delta + \epsilon + \zeta}{2}\\
&\mathrm{Cov}[X,Y] = \frac{\delta + \epsilon - \zeta}{2}\\
\end{aligned}
\end{equation*}

Therefore, for a given cortical morphology measure, say cortical thickness, we can infer its error variance and covariance, and hence we can visualise the error structure with an ellipse (see section \ref{ellipsemethod}). We show this in Fig.~\ref{SingleCorrError}(B) and (C) in the middle column. From Var(X), Var(Y), and Cov(X,Y) we can then estimate the `true' underlying correlation in the absence of error. This is displayed in the right column in Fig.~\ref{SingleCorrError}(B) and (C).

We will also use the term `attenuation', by which we refer to the difference between the measured correlations and the estimated true correlation. Usually the latter is larger than the former, hence the term attenuation of correlation.

\subsubsection{Simulating the effect of measurement error}

For our simulation in Fig.~\ref{AttenuationReliability}(A), we artificially generated data to demonstrate the effect of measurement error. We computed two sets of 1000 random data points sampled from a bivariate normal distribution with zero mean and unit variance. These data vectors represent the ground truth morphological data of two ROIs. We can generate them with any pre-defined correlation of the ROIs, which we use as the true underlying correlation. 

We then added a random normal error vector (zero mean, and $\nu$ standard deviation, where $\nu$ represents the relative error/noise magnitude to the measurement) to each of the ROIs to simulate noisy measurements of the morphological data. The correlation between these simulated measurements corresponds to the measured correlations in the real data. We repeated this process 10000 times to compute the variability of the simulated measured correlations. 


As a measure of `unreliability' we computed the standard deviation over the 10000 simulated measured correlations, where a high standard deviation (high `unreliability') indicates low reliability. Further, we computed the difference between the true correlation and the mean correlation over the 10000 simulation runs as the attenuation.

We simulated two example levels of true correlation (0.7 and 0.4) each with two different example levels of relative noise (25\% and 50\%) in Fig.~\ref{AttenuationReliability}.

\subsection{Data availability statement}
Analysis code and data of all morphological measures (selected subjects of HCP, HCP scan and rescan, selected subjects of Cam-CAN preprocessed in all FreeSurfer versions) are available at Github (\url{https://github.com/cnnp-lab/2019Carmon-ReliabityComparabilityStructuralCovariance}).

\section{Results}

\subsection{Differences in structural covariance of cortical thickness between different sites, scan-rescan, and different FreeSurfer versions.}

We first focus on structural covariance of cortical thickness, which is widely used morphology measure. To investigate the comparability of structural covariance we compared two data sets of healthy human subjects (HCP and Cam-CAN data restricted to the same number of males and females within a narrow age range of 22-34 year old). We standardised the data for each site in each ROI before comparison (effectively applying a site correction to each ROI). 

\begin{figure}[H]
  \centering
  \vspace*{-3.85cm}
  \hspace*{-3.1cm}
  \includegraphics[width=180mm]{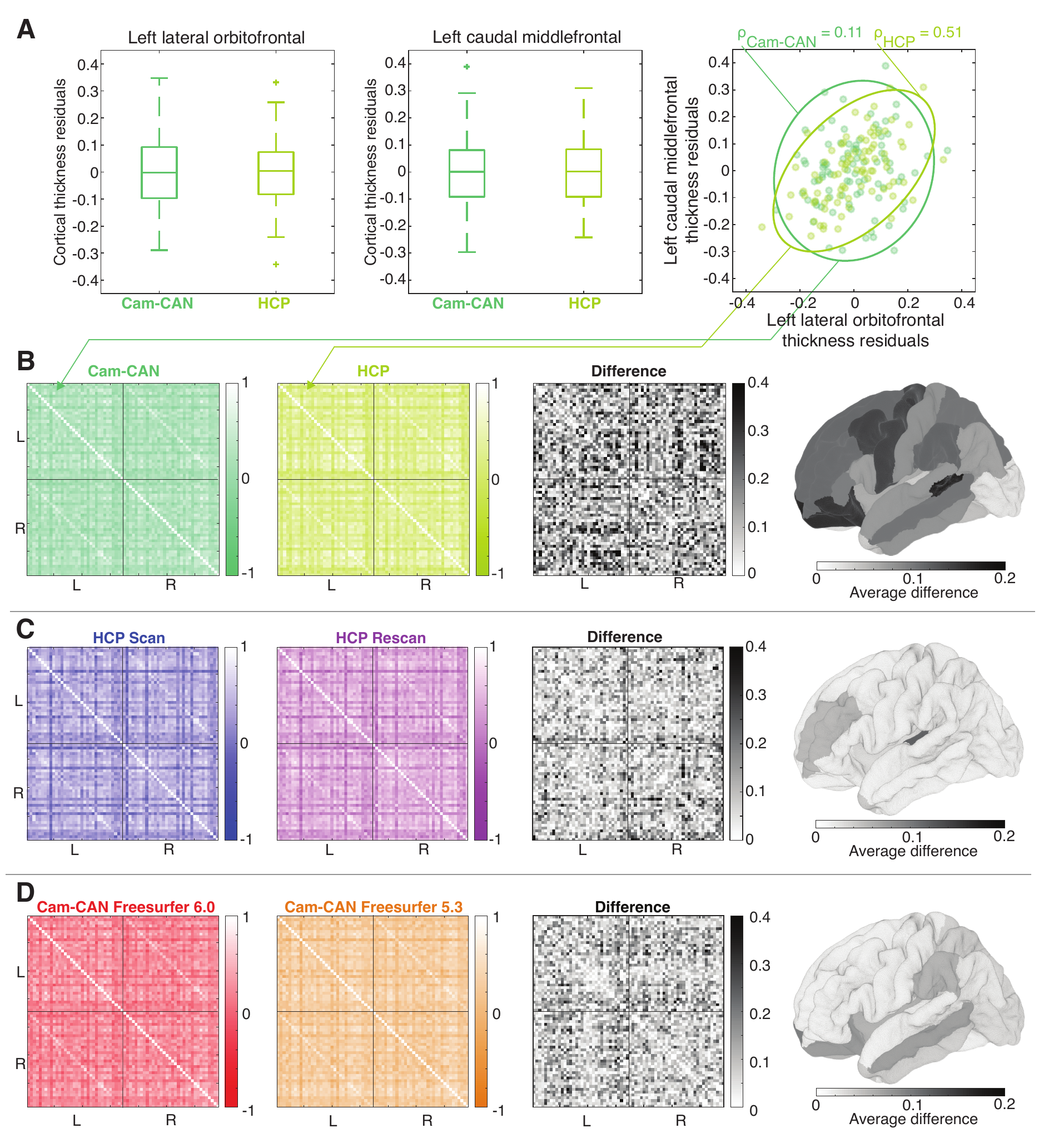}
  \vspace*{-0.5cm}
  \caption{\textbf{\textbf{Comparability and reliability of cortical thickness structural covariance.}} (A) After standardisation ($\mu = 0$, $\sigma = 1$), two example ROIs are comparable in their cortical thickness distribution between Cam-CAN (mint green) vs. HCP (lime green). Nevertheless, the correlation of the ROIs still differ significantly between the two data sets (right panel, p = 0.003 **). Ellipsoids in the right panel visualise the correlation structure. (B) The first two columns show the full structural covariance matrix based on the Desikan-Killiany atlas. The third column displays the absolute difference between these structural covariance matrices in each matrix entry. The last column shows the average difference on the brain as a heatmap. (C) Same as (B) but for HCP scan data (blue) and rescan data (purple). (D) Also same as (B) but for Cam-CAN processed in FreeSurfer version 6.0 (red) and FreeSurfer version 5.3 (orange). Note that the colour code used for each data set will be maintained throughout the remaining figures.}
  \label{CorrDiffAll}
\end{figure}

In Fig.~\ref{CorrDiffAll}(A) we see that after site-correction the thickness of two example ROIs are comparable (same mean and variance) between HCP and Cam-CAN. However, the correlation between the ROIs remains significantly different between the sites (p=0.003 permutation test). We demonstrate with this example that, as mathematically expected, site correction of the univariate measures does not correct for differences in correlation. 

In Fig.~\ref{CorrDiffAll}(B) we show the raw structural covariance (i.e. correlation matrix) of each dataset. We also show the absolute differences between the HCP and Cam-CAN  correlation matrices. Mean differences for each ROI are also visualised as a heatmap on the brain. In the Supplementary Material S16 and S17, we show additional site comparisons using other datasets, demonstrating differences in the structural covariance matrix that are in a similar range as shown here.

To investigate the reliability of scan sessions on the same subjects, we used the HCP scan-rescan data. Similar to Fig.~\ref{CorrDiffAll}(B), we show the raw structural covariance matrices and their differences between the sessions in Fig.~\ref{CorrDiffAll}(C). Generally, a weaker difference is seen compared to the site comparison. 
In the Supplementary Material S14, we additionally show  results for another scan-rescan dataset, where the range of differences in the structural covariance matrix is slightly higher.

Finally, we show the effect of using different FreeSurfer versions of the same subjects in Fig.~\ref{CorrDiffAll}(D). The results show again that there are substantial differences in correlation. Albeit, they are overall less pronounced than in the site comparison. 

In summary, we detected (significant) differences of structural covariance between two demographically comparable data sets of healthy human subjects. These differences are prominent despite site correction. Similarly, the structural covariance across scan sessions and FreeSurfer versions (Fig.~\ref{CorrDiffAll}(C) and (D)) also show differences, which are less pronounced in effect than between sites.

\newpage

 \subsection{Significant differences in structural covariance of cortical thickness between sites}
 
 \begin{figure}
  \hspace{-1.5cm}
  \includegraphics[width=160mm]{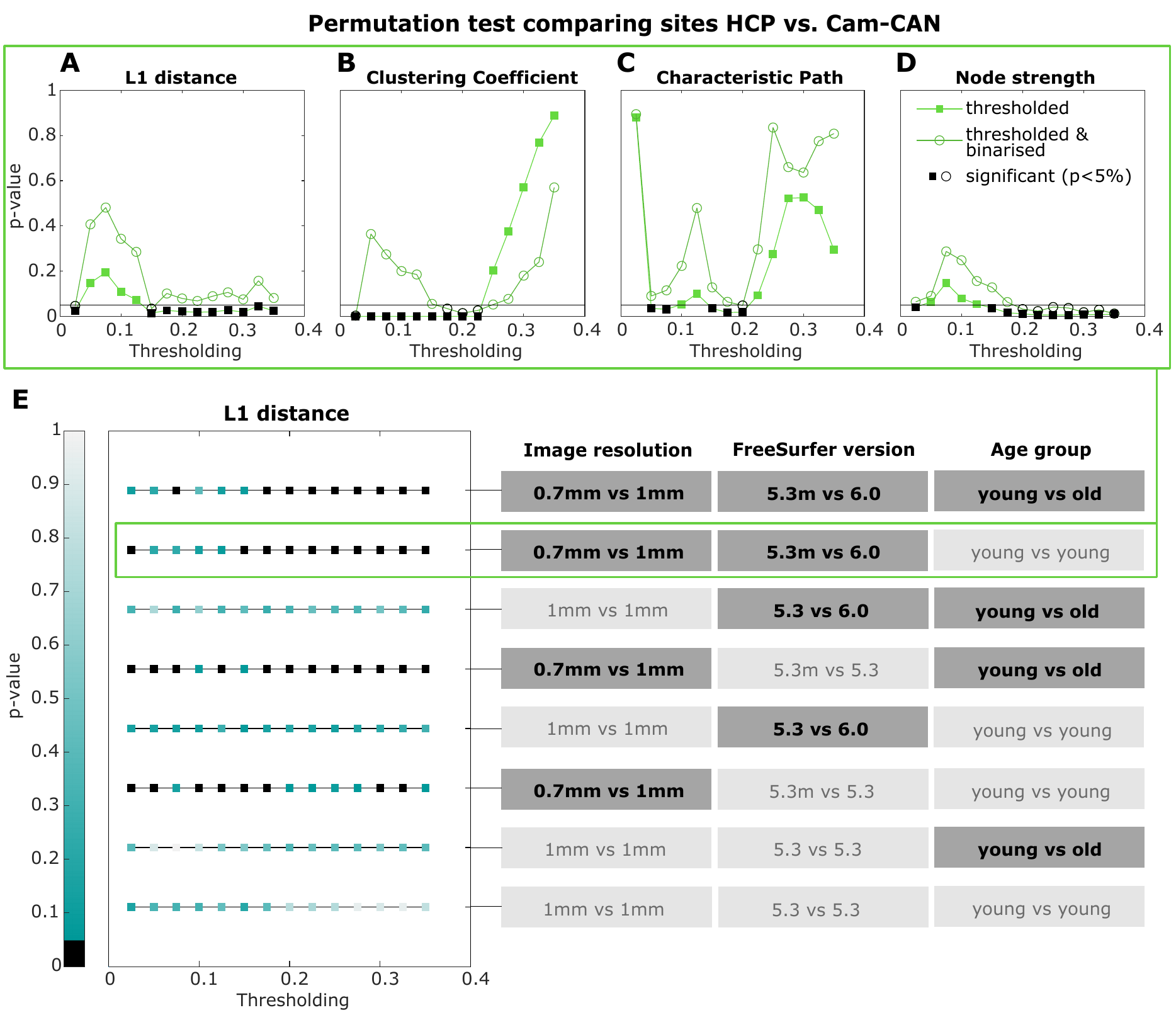}
  \caption{\textbf{Significant differences in network measures when comparing healthy subjects from different sites.} (A) shows the p-values of the $L_{1}$ distance in the comparison between HCP and Cam-CAN. (B,C,D) show p-values based on other common network measures. Solid lines correspond to the thresholded matrix comparisons and dashed lines correspond to the thresholded and binarised matrix comparisons. The y-axis indicates the p-value of the comparison. Black markers indicate a $p<0.05$ for reference. (E) depicts the p-values of the thresholded $L_{1}$ distance for different versions of HCP and Cam-CAN. The difference in the versions is summarised as combinations of differences in imaging resolution, FreeSurfer version and age group on the right. Image resolution is shown in isotropic voxel size. FreeSurfer 5.3m is the modified FreeSurfer 5.3 version, used by the HCP preprocessing for the 0.7mm resolution. The young (old) age group were subjects in the age range of 22-34 (55-67).}
  \label{NetworkDiff}
\end{figure}
 
In Fig.~\ref{CorrDiffAll}(B) we showed that the structural covariance (correlation matrix) shows substantial differences between two sites. Here, we investigate if these differences also imply significant differences for graph theoretic network measures under a permutation test (Fig.~\ref{PermutationTest}).  From the thresholded (and binarised) correlation matrices we computed the $L_1$ distance, node strength/degree, characteristic path and clustering coefficient. Fig.~\ref{NetworkDiff}(A-D) shows the p-values of the comparison between sites for different thresholds (ranging from 2.5\% to 35\%). Significant differences ($p<0.05$) are seen in all network measures in both the thresholded and binarised cases for a range of thresholds. In supplementary Fig. S1 we also compare additional network measures. 

In the site comparison so far, we used two sites that differ slightly in a range of factor. Other than using different subjects (of comparable demographics), the image resolution, and FreeSurfer version used for processing the two different sites also differed. To  investigate the influence of these factors on producing significant network differences between sites, we systematically varied these factors in different combinations (Fig.~\ref{NetworkDiff}(E)). For example we can make the site comparison with different image resolutions, FreeSurfer versions, and using different subject demographics (young vs. old in this case) -- this is depicted in the first row of Fig.~\ref{NetworkDiff}(E).  We can also make the site comparison with the same image resolution, and comparable subject demographics, but only with different FreeSurfer versions (shown in the 5th row of Fig.~\ref{NetworkDiff}(E)). Finally, the site comparison can also be made using the same resolution, the same FreeSurfer version, and the same subject demographics (last row of Fig.~\ref{NetworkDiff}(E))). We also show the corresponding p-values across different thresholds for each of these scenarios, using L1 distance as a network measure. 

Fig.~\ref{NetworkDiff}(E) shows that the more factors differ between the site comparisons, generally, the more we tend to observe significant ($p<0.05$) differences in L1 distance. The difference in imaging resolution between sites appears to be best correlated with significant differences in L1 distance. However, we note that the FreeSurfer version used for the HCP 0.7mm isotropic resolution is a modified version of FreeSurfer version 5.3. Therefore the pure effect of imaging resolution when compared to 1mm isotropic Cam-CAN (FreeSurfer version 5.3) cannot be ascertained.  

Our supplementary materials S15 offers further analyses on the effect of image resolution, demonstrating that decreasing image resolution leads to further differences in the structural covariance matrix. Supplementary materials S16 and S17 show additional site comparisons with other sites/datasets, using a wider age-range of subjects and different FreeSurfer versions. Those analyses show that substantial differences are also seen between other sites, and some significant differences are observed in these datasets particularly in the clustering coefficient.
For simplicity and consistency with previous figures, we will continue to use the setting in the second row of Fig.~\ref{NetworkDiff}(E) for our subsequent analyses of site comparisons. 
 
In summary, we show that the substantial differences in the structural covariance matrix between HCP and CamCAN sites translate into significant differences of network measures under a permutation test. However, by carefully matching the sites in different factors (here: image resolution, FreeSurfer version, and demographics) the differences are no longer statistically significant in our comparison.

\subsection{Average cortical thickness shows larger differences in structural covariance than surface area and volume}

\begin{figure}
  \centering
  \includegraphics[width=90mm]{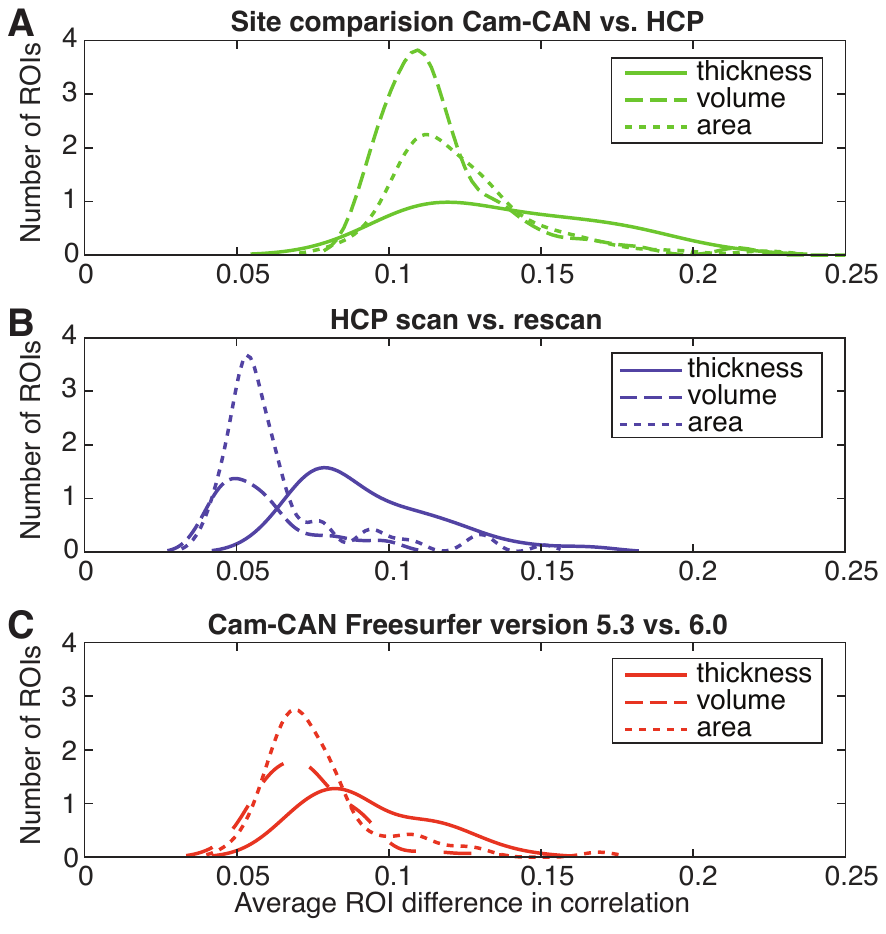}
  \caption{\textbf{Density plot of average difference in each ROI between sites, scan sessions, and FreeSurfer versions.} (A) shows differences between site (HCP and Cam-CAN), (B) shows differences between the HCP scan and rescan data, and (C) displays the differences between FreeSurfer versions. Solid lines correspond to average cortical thickness, dashed lines to volume and dotted lines to surface area. We calculated the average difference by computing the ROI-wise mean of the absolute difference between the respective structural covariance matrices. Densities were estimated with a nonparametric kernel-smoothing method using 100 equally spaced points.}
  \label{MorphologyMeasureDiff}
\end{figure}

Average cortical thickness, surface area, and volume are three of the most popular measures to analyse structural covariance. Thus far we only investigated cortical thickness derived networks. Next, we investigated if the structural covariance of other morphological measures are more comparable and reliable. 

Fig.~\ref{MorphologyMeasureDiff} shows the estimated distribution density of average differences in each ROI in their structural covariance matrix using different measures of cortical morphology. The different panels display the differences between site, scan/rescan, and FreeSurfer versions, respectively. In all panels the differences are most pronounced for average cortical thickness, where more ROIs display larger differences compared to the other cortical morphology measures. We can also note, as in Fig.~\ref{CorrDiffAll}, that generally the differences between sites are larger than the differences between scan sessions and FreeSurfer versions.
In the supplementary material S13 we also include corresponding brain surface plots (as in Fig. \ref{CorrDiffAll}) for all morphological measures. 

In summary, the choice of the morphological measure also affects the comparability and reliability of the structural covariance. In our study average cortical thickness is associated with larger differences than surface area and volume when comparing sites, scan/rescan, and FreeSurfer versions. 

\subsection{Lowest correlations, largest estimated error and strongest estimated attenuation in average cortical thickness}

To investigate the source of differences in correlation we used a single correlation as an initial starting point: namely the correlation between the left and right hemispheres. On the level of a single correlation we can visualise important aspects, and show why cortical thickness shows the prominent differences we found in Fig.~\ref{MorphologyMeasureDiff} as opposed to surface area and volume. 

\begin{figure}[H]
\vspace{-0.5cm}
\hspace*{-3cm}
  \includegraphics[width=180mm]{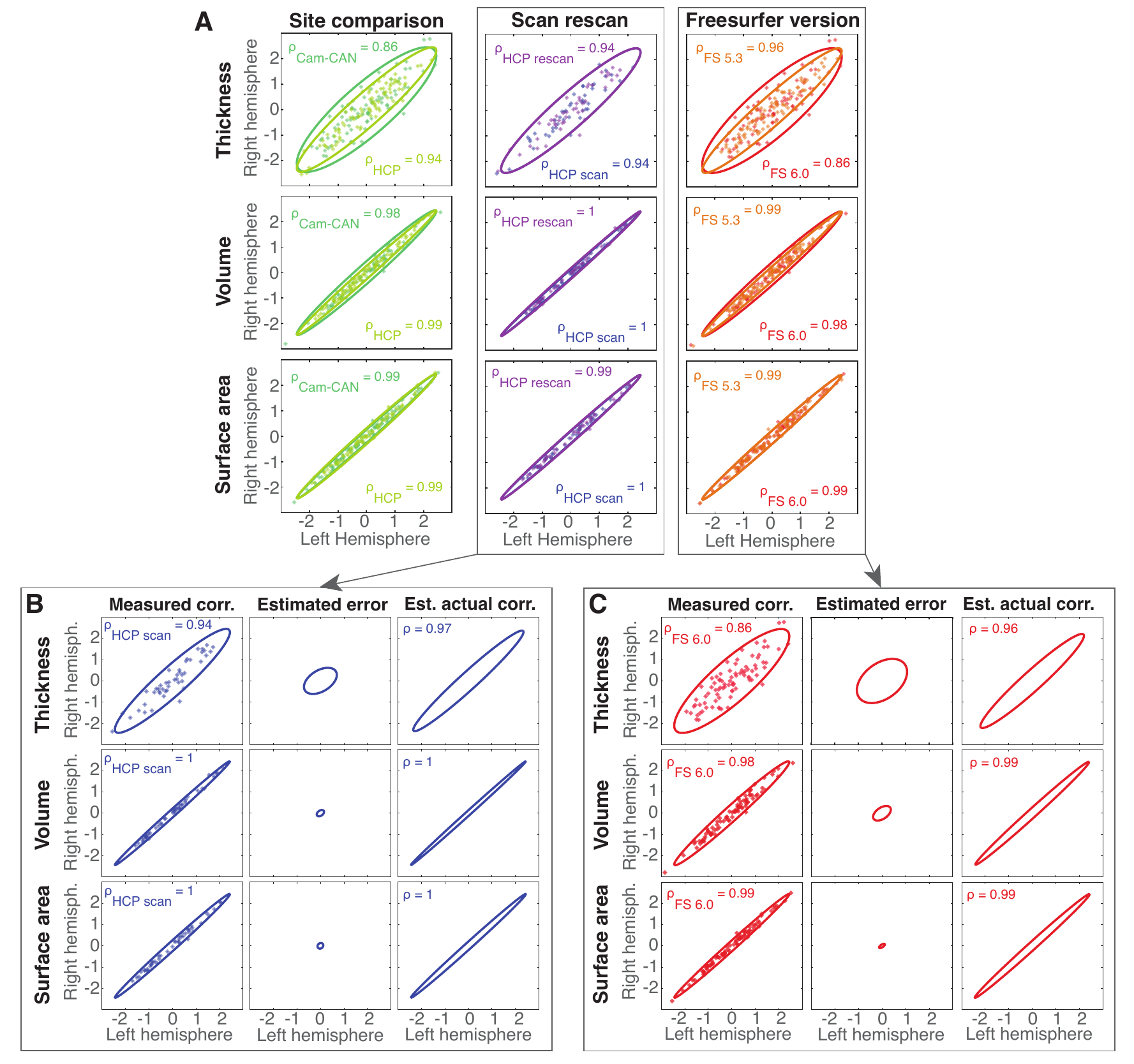}
  \caption{\textbf{Correlation of the left and the right hemisphere and their estimated errors.} (A) Scatter plot of the left vs the right hemisphere for different morphological measures. Correlations of HCP and Cam-CAN site comparison are shown in the left column. Comparison of scan/rescan data and FreeSurfer versions are displayed in the middle and right column. To better visualise each correlation we show an ellipse. (B and C) For subjects with repeat measurements, we could estimate the underlying relative error variance and covariance, and `true correlation'. In both panels the left column shows the measured correlation, the middle column the estimated relative error and the right column the estimated underlying `actual' correlation.}
  \label{SingleCorrError}
\end{figure}

Fig.~\ref{SingleCorrError}(A) shows the scatter plot of the left vs right hemisphere. Each dot corresponds to a subject in one of the data sets. Each row depicts one morphological measure and each column one data pair comparison (HCP/CamCAN, scan/rescan, and FreeSurfer version 5.3/6.0). As to be expected from Fig.~\ref{MorphologyMeasureDiff} we see that the difference between the data sets are largest in the correlations of average cortical thickness. We also observe that the correlations of average cortical thickness are lowest in each data set. This finding also extends to the correlations of ROIs of the Desikan Killiany atlas (see supplementary Fig. S2). 

For the scan/rescan data and the FreeSurfer 5.3/6.0 datasets it is possible to estimate 1) the measurement error and 2) the underlying correlation. The rationale behind this estimation is to utilise the ``repeated measurements'' from the same subject, either in scan/rescan, or in two different FreeSurfer versions. From these repeated measurements we analytically estimated the variance and covariance of the measurement error as well as the underlying ``true'' correlation (Fig.~\ref{SingleCorrError}(B) and (C)) - see section \ref{esterrorsection}. We observe that the estimated error is largest for average cortical thickness and smallest for surface area. Fig.~\ref{SingleCorrError}(B) and (C) also shows that all measured correlations were attenuated by the error structure, such that the estimated underlying correlation is always larger than the empirically measured correlation. We also remark that the error introduced by the FreeSurfer versions is comparable in magnitude to that of scan/rescan. Note that the error estimated here is relative to unit variance of the measurement (see Methods section \ref{sectionstats}), such that all the panels in Fig.~\ref{SingleCorrError} can be directly compared to each other. 

The findings of Fig.~\ref{SingleCorrError}(B) and (C) also extend to most ROI pairs of the Desikan Killiany atlas (see supplementary Fig. S3). 

In summary, we observed that average cortical thickness is least correlated, the error of average cortical thickness is largest, and the estimated underlying correlations of average cortical thickness are most attenuated.
 
\newpage

\subsection{A larger error leads to greater levels of attenuation and unreliability}

\begin{figure}
\vspace*{-2cm}
\hspace*{-1cm}
  \centering
  \includegraphics[width=140mm]{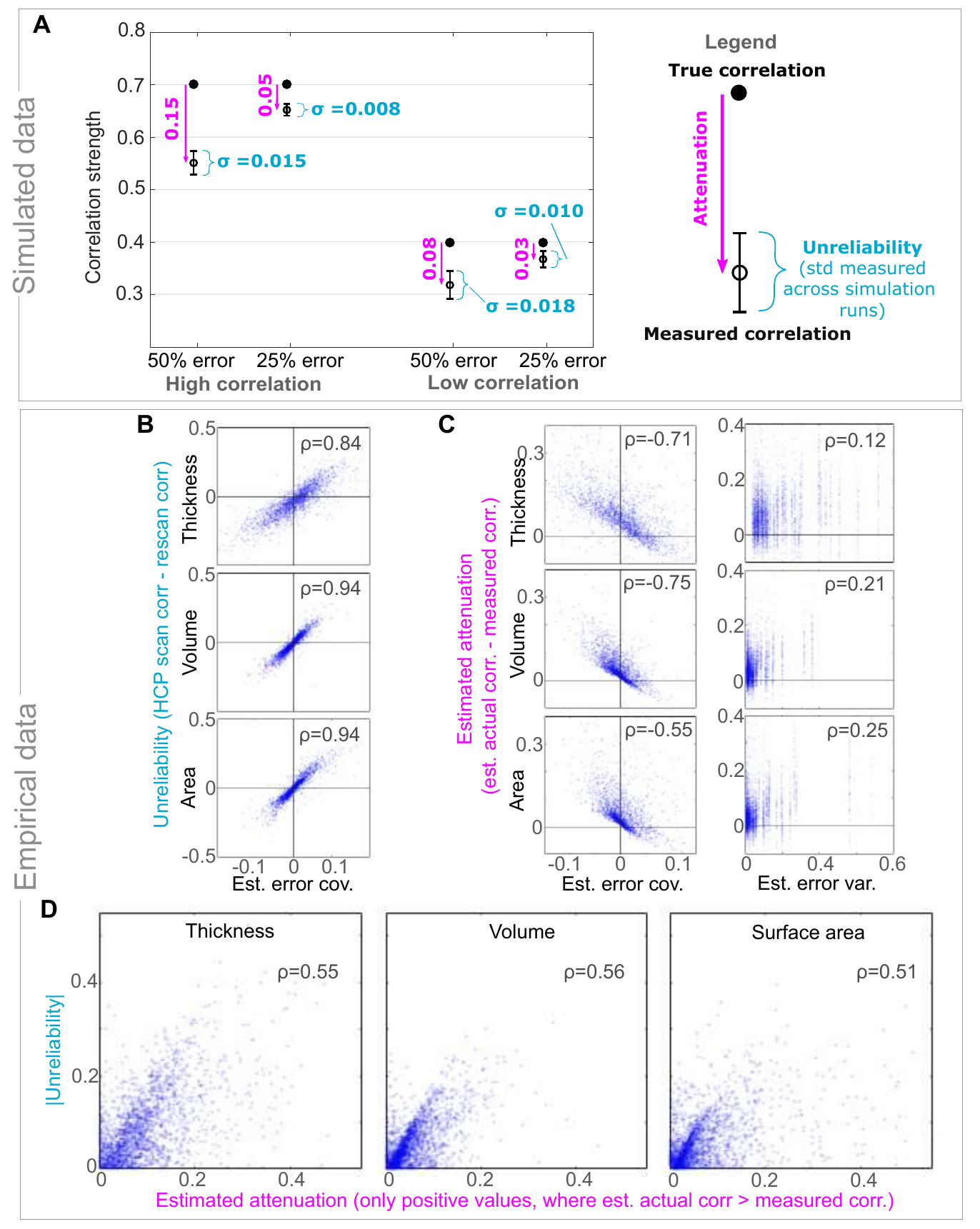}
  \vspace*{-0.2cm}
  \caption{\textbf{A larger error leads to greater levels of attenuation and unreliability.} (A) Simulated data with two different levels of correlation and error. The legend on the right provides the key to interpreting the schematic illustration of the effect of error on simulated correlations. (B)-(D) uses the HCP scan data (hence blue data points). Each data point is a ROI-pair from the structural covariance matrix. (B) shows the estimated error covariance scattered against the measured correlations of HCP scan minus measured correlations of HCP rescan (which we term Unreliability). (C) shows the estimated covariance error (left column) and estimated variance error (right column) scattered against the estimated `corrected' correlation minus the measured correlation of HCP scan, i.e. the estimated attenuation. (D) displays the estimated attenuation scattered against the absolute unreliability.}
  \label{AttenuationReliability}
\end{figure}

We demonstrated so far that thickness is less reliable than volume and surface area, and we estimated a larger error structure and a stronger attenuation of correlations for thickness. Here, we relate these observations to each other using the full Desikan-Killiany ROI correlation matrices. 

In Fig.~\ref{AttenuationReliability}(A) we investigated the effect of error on reliability and attenuation for simulated data. Using simulated data with predefined true correlation levels (0.7 and 0.4), we added relative error of two different magnitudes (25\% and 50\% error). We repeated the simulation 10,000 times to compute the variation of the measured correlations. A smaller variation between the different simulations suggests greater reliability, and here we define the variance/difference between simulations as a measure of `unreliability'. In supplementary Fig. S8 we show that, in our simulations, the true correlations and the strength of the added error are well-captured by our estimations.


In the simulations, we note two key observations. First, stronger errors increase unreliability and proportionally increase the attenuation, regardless of the strength of true correlation. Second, for the same error magnitude, high correlations are more attenuated than low correlations. The correlation strength minimally affects the effect of error on unreliability. In empirical data, and if our estimates of error are correct, we would thus expect to see error variance and covariance to be associated with the attenuation and the level of unreliability. In turn, we expect an association between unreliability and attenuation to arise. This is indeed observed in Fig.~\ref{AttenuationReliability}(B)-(D), where we used all entries of the Desikan-Killiany atlas matrix to demonstrate these associations.

Fig.~\ref{AttenuationReliability}(B) shows a very high association between the estimated error covariance of the HCP scan data and the level of unreliability (difference between the measured correlations of HCP scan and rescan). In absolute terms, correlations with a stronger absolute error covariance show more difference between scan/rescan. This agrees with our simulations, where a larger error magnitude leads to a higher level of unreliability. Note that unlike the estimated error covariance, the estimated error variance is the same between different measurements for standardised data (since $1=Var[X] = Var[Y]$, $\alpha$, $\beta$ and $\gamma$ of section \ref{esterrorsection} are the same for $X$ and $Y$) and therefore does not contribute to the high unreliability. In simple words, the larger the estimated error covariance, the more the correlation coefficients differ between scan and rescan.

In Fig.~\ref{AttenuationReliability}(C), left column, we can see the negative associations between the estimated attenuation and the estimated error covariance. More negative covariances correspond to a larger decrease in the measured correlation compared to the estimated actual correlation, and \textit{vice versa}. For the error variance, a larger error variance corresponds to a larger decrease in the measured correlation compared to the estimated actual correlation. In summary, more attenuated correlations are associated with stronger absolute error (co)variance, in agreement with our simulation. 


%

Most correlations are positively attenuated ($corrected$ $correlation$ $-$ $measured$ $correlation$ $> 0$). In Fig.~\ref{AttenuationReliability}(D) we excluded the small proportion of correlations that show a negative attenuation. The plot associates more attenuated correlations with a stronger absolute unreliability (higher absolute differences between the measured correlations of scan/rescan). In other words, more attenuated correlations tend to be less reliable (bigger difference between scan and rescan). This agrees with our observation in the simulated data that reliability is proportional to attenuation.  

In summary, we showed on simulated data that a stronger error causes a stronger attenuation of the correlation and increases unreliability proportionally. The empirical data supported this finding. We can thus link the two observations of attenuation and unreliability: more strongly attenuated correlations tend to be less reliable.

\section{Discussion}

To our knowledge, this is the first study to investigate the comparability (different subjects) and reliability (same subjects) of structural covariance analysis.  We showed that site differences in structural covariance are naturally not accounted for with site correction of the univariate brain morphology measures. Common down-stream analyses of different network measures also show significant differences between sites. In terms of reliability, scan vs. rescan data and the same images processed by different FreeSurfer versions also show  differences in their structural covariance, albeit less prounounced in magnitude compared to site-comparisons. Interestingly, we also found that the magnitude of the differences vary between morphological measures. In our analysis, the structural covariance of cortical thickness is least reliable and comparable. We further showed that the estimated relative error is largest in thickness, where the estimated attenuation of the correlation is also the strongest. Finally, by using simulated data in combination with empirical observations, we argue that measurement errors attenuate correlations and cause a low reliability, which is particularly pronounced in cortical thickness.

In addition to FreeSurfer version and scan session, difference in image resolution also drive differences in structural covariance. To reduce the complexity of the paper, we presented those investigations in Supplementary Materials S15. We show that the differences in structural covariance increase dramatically as image resolution is lowered, particularly in cortical thickness. In agreement with this observation, we could also see in Fig. \ref{NetworkDiff}(E) that differences in image resolution is associated with significant differences between sites. This is further investigated by additional site comparisons in Supplemental Materials S16 and S17.  

Commonly the structural covariance is computed with Pearson's correlation coefficient, to which also our main results are restricted to. An alternative correlation coefficient is the Spearman's rank correlation, known to be less prone to outliers. However, we do not find major changes to our results using Spearman's correlation (see supplementary Fig. S11). An important question to address is if there are other correlation types which are less prone to errors in covariance structure. For example, Geerligs \textit{et al.} show in their study that multivariate distance correlation is more reliable than Pearson's correlation under certain circumstances \cite{geerligs2016functional}. Structural covariance is also measured in other ways e.g. partial least squares regression \cite{spreng2019structural}. Future studies should investigate the comparability and reliability of different measures of structural covariance.     

In this context of covariance estimation, it is also important to acknowledge the importance of sample size. The estimation of covariance is inherently noisy, especially with low sample sizes. We investigated the comparability aspect in that regard in Supplementary Materials S18. We found that the differences in the structural covariance matrices between distinct subsets of the CamCAN data do not decrease substantially beyond 70-80 subjects (Fig. S18). However, for less than \~30 subjects the differences increase rapidly. Similar behaviour -increasing differences with decreasing sample size- was observed in our scan-rescan data in terms of reliability (Supplementary Materials S19). We thus recommend using as large a sample size as possible, ideally n>>30 for the Desikan-Killiany parcellation. Although some existing studies use relatively large sample sizes, many of those are pooled data sets across different scanners and sites, which may not be comparable in terms of their structural covariance. Large and homogeneous datasets from a single site are still rare. This essentially means that estimators of covariance that perform well with low sample sizes are needed. Future studies could investigate the effect of e.g. shrinkage estimators \cite{rahim2019population} on the reliability and comparability of structural covariance.

One way of improving reliability may be achieved by estimating the true underlying correlation from repeated measurements. Our approach could be used for such a purpose, and in artificial data we could show that the corrected correlations are close to the true underlying correlations (see supplementary Fig. S9). 
However, we had to make several assumptions to arrive at the estimated corrected correlation. Future work should investigate the validity of these assumptions further in specific contexts such as scan and rescan. Especially where there are several repeated measurements the true correlation can be estimated with other frameworks (e.g. Bayesian \cite{behseta2009bayesian}). Importantly, if the error covariance structure is derived once, then it can be applied to correct structural covariance matrices, even across scanners and sites. Theoretically, all that is required is that a group of subjects are scanned on both scanners/sites to infer the error covariance structure. Once inferred, the correction can be applied to any future subject groups to make them comparable across sites. 

Additionally, reliability could also be improved by the choice of ROIs.  Previous research showed that different ROIs are differently reliable in their univariate cortical morphology measures \cite{han2006reliability}. This may also extend to their covariance structure. There are many potential criteria in restricting ROIs which could improve reliability and comparability. One criterion is to select ROIs by their size. Indeed, we could find a small effect of the ROI volume, where smaller ROIs tended to have lower p-values under a permutation test (see supplementary Fig. S10). ROIs could also be selected by their spatial proximity (e.g. movement artifacts would cause covarying errors between ROIs). Relatedly, voxel-based analyses of structural covariance (e.g. \cite{mechelli2005structural}) is a popular alternative approach to the ROI-based approach presented here. We expect measurement error to still affect the reliability and comparability of voxel-based structural covariance. However, it is unclear if some processing steps (such as spatial warping and smoothing) used in voxel-based approaches exacerbate or alleviate the covarying measurement errors. A future study comparing the two approaches in terms of their reliability and comparability will be required. 

Previously, studies reported distinctions between the structural covariance of different cortical measures \cite{sanabria2010surface,yang2016complementary}. Yang \textit{et al.} found statistically significant differences between the structural covariance of cortical thickness, volume and surface area \cite{yang2016complementary}. Although that observation was not made in the context of reliability and comparability, it agrees with our observation that there are some inherent differences between thickness and surface area covariances. It is alarming that in our study the structural covariance of average cortical thickness is the least reliable and comparable of the cortical morphology measures, as it is the most commonly used measure for structural covariance. It is known that cortical thickness has a lower level of biological variance compared to surface area and volume (also see supplementary Fig. S12). In relative terms, the same magnitude of error would thus have a stronger effect. In agreement, we indeed show a higher relative error for cortical thickness compared to surface area, or volume.

The results of our study are of particular interest for the comparison of different clinical conditions, as some studies combine sites to increase their sample size. In our investigation we could see significant differences between sites, even when using comparable healthy subjects with similar demographics. Thus, ideally for the comparison of clinical conditions we recommend performing the analysis for each site separately, and only test for agreement across sites. Advanced hierarchical modelling may help in estimating a more reliable and comparable covariance structure across sites in future work. This might be of particular interest for ongoing studies, especially if thickness has been used as cortical measure. 

From a biological perspective, the argument for analysing structural covariance is that a strong covariance indicates co-regulation, or co-development \cite{mechelli2005structural}. It is however unclear if simply obtaining the covariance of a cortical measure across several ROIs is indeed the best way to capture and analyse such hypothesised co-development, or co-regulation patterns. For example, if there is no biological co-relation between two ROIs, then all that will be measured in experiments is the error and correlation in the errors. The study of covariance is a way of dimensionality reduction, and a more robust method of dimensionality reduction inherently utilizing reliability (see e.g. Sotiras \textit{et al.} \cite{sotiras2017patterns} ) may be better suited for a reliable and comparable data-driven approach.

Alternatively, hypothesis-driven approaches can be taken, where a specific covariance between specific morphological variables and/or ROIs is predicted by theory. If the theory is correct, the data should support the theory in a comparable and reliable manner. One example of such a predicted covariance structure is the recently-described `universal scaling law of cortical folding' \cite{wang2019human, wang2016universality, mota2015cortical}. According to this law, brain surface area, cortical thickness, and exposed surface area are linked by a strong covariance structure, which has been confirmed across species\cite{mota2015cortical}, within human populations\cite{wang2016universality}, and even between different ROIs of the same brain\cite{wang2019human}. Importantly, the scaling law has been shown to be comparable across sites and reliable within the same data set. Future work could take a combined data-driven and hypothesis-driven approach to discover true covariance structures in brain morphology that are reliable and comparable.

In summary, our analyses show that the question of comparability and reliability is crucial in the study and interpretation of structural covariance. Reliable or comparable univariate measures of ROIs do not imply reliable and comparable correlations between them. Practically, we make the following recommendations (1) combining subjects across sites into one group for structural covariance analysis should be avoided, particularly if sites differ in image resolutions, subject demographics, or preprocessing steps; (2) surface area and volume should be preferred as morphological measures of structural covariance over cortical thickness; (3) an analysis of required sample size should be performed, ususally the larger the sample size, the better the correlation estimate. We recommend to  use n>>30 subjects per group for the Desikan-Killiany parcellation; (4) measurement error should be assessed where repeated measurements (e.g. from scan-rescan) are available; (5) if combining sites is absolutely required, univariate (per ROI) site-correction is insufficient, but error covariance (between ROIs) should be explicitly modelled. Future work should establish a pipeline of reliable and comparable structural covariance analysis based on robust data-driven dimensionality reduction and hypothesis-driven discovery of existing covariance structures.

 \section*{Acknowledgement}
 Data were provided in part by the Human Connectome Project, WU-Minn Consortium (Principal Investigators: David Van Essen and Kamil Ugurbil; 1U54MH091657) funded by the 16 NIH Institutes and Centers that support the NIH Blueprint for Neuroscience Research; and by the McDonnell Center for Systems Neuroscience at Washington University. 
Another part of the data for this project was provided by the Cambridge Centre for Ageing and Neuroscience (CamCAN). CamCAN funding was provided by the UK Biotechnology and Biological Sciences Research Council (grant number BB/H008217/1), together with support from the UK Medical Research Council and University of Cambridge, UK.
 
We thank members of the CNNP lab (\url{www.cnnp-lab.com}) for discussions on the analysis and manuscript. P.N.T. and Y.W. gratefully acknowledge funding from Wellcome Trust (208940/Z/17/Z and 210109/Z/18/Z). J.H.N. was supported by the Reece Foundation. M.K. was supported by Wellcome Trust (102037), Engineering and Physical Sciences Research Council (NS/A000026/1, EP/N031962/1), Medical Research Council (MR/T004347/1), and the Guangci Professorship Program of Ruijin Hospital (Shanghai Jiao Tong Univ.). The authors declare no conflict of interest. The funders played no role in the design of the study.

\section*{Author contributions}
J.C. and Y.W. conceived the idea, developed the methods, wrote the code, and performed the analysis. J.H. contributed ideas on statistical analyses. J.H.N. helped with FreeSurfer processing of the data. J.C. and Y.W. drafted the manuscript and produced all figures. All authors participated in critically reviewing and revising the manuscript.

\bibliography{references.bib}

\end{document}